\documentclass[
aip,
 amsmath,amssymb,
 reprint,%
]{revtex4-1}

\usepackage{graphicx}
\usepackage{dcolumn}
\usepackage{bm}
\usepackage[normalem]{ulem}

\usepackage[utf8]{inputenc}
\usepackage[T1]{fontenc}
\usepackage{mathptmx}
\usepackage{etoolbox}
\usepackage{xcolor}
\usepackage{filecontents}

\definecolor{applegreen}{rgb}{0.55, 0.71, 0.0}

\makeatletter
\def\@email#1#2{%
 \endgroup
 \patchcmd{\titleblock@produce}
  {\frontmatter@RRAPformat}
  {\frontmatter@RRAPformat{\produce@RRAP{*#1\href{mailto:#2}{#2}}}\frontmatter@RRAPformat}
  {}{}
}%
\makeatother
\begin{document}

\preprint{AIP/123-QED}

\title{Quantum-interference origin and magnitude of 1/$f$ noise in Dirac nodal line IrO$_2$ nanowires at low temperatures}
\author{Po-Yu Chien}

\affiliation{ 
Department of Electrophysics, National Yang Ming Chiao Tung University, Hsinchu 30010, Taiwan
}
 
\author{Chih-Yuan Wu}%
 

\affiliation{ 
Department of Physics, Fu Jen Catholic University, Taipei 24205, Taiwan
}%

\author{Ruey-Tay Wang}

\affiliation{ 
Department of Electrophysics, National Yang Ming Chiao Tung University, Hsinchu 30010, Taiwan
}

\author{Shao-Pin Chiu}
\affiliation{ 
Department of Electrophysics, National Yang Ming Chiao Tung University, Hsinchu 30010, Taiwan
}

\author{Stefan Kirchner}
\affiliation{ 
Department of Electrophysics, National Yang Ming Chiao Tung University, Hsinchu 30010, Taiwan
}
\affiliation{ 
Center for Emergent Functional Matter Science, National Yang Ming Chiao Tung University, Hsinchu 30010, Taiwan
}

\author{Sheng-Shiuan Yeh}
\altaffiliation [Authors to whom correspondence should be addressed: ] {ssyeh@nycu.edu.tw (S.S.Y.) and jjlin@nycu.edu.tw (J.J.L.)}

\affiliation{ 
Center for Emergent Functional Matter Science, National Yang Ming Chiao Tung University, Hsinchu 30010, Taiwan
}

\affiliation{ 
International College of Semiconductor Technology, National Yang Ming Chiao Tung University, Hsinchu 30010, Taiwan
}

\author{Juhn-Jong Lin}
\altaffiliation [Authors to whom correspondence should be addressed: ] {ssyeh@nycu.edu.tw (S.S.Y.) and jjlin@nycu.edu.tw (J.J.L.)}

\affiliation{ 
Department of Electrophysics, National Yang Ming Chiao Tung University, Hsinchu 30010, Taiwan
}

\date{\today}

\begin{abstract}
We present 1/$f$ noise measurements of IrO$_2$ nanowires from 1.7 to 350 K. Results reveal that the noise magnitude (represented by Hooge parameter $\gamma$) increases at low temperatures, indicating low-frequency resistance noise from universal conductance fluctuations. The cause of this noise is determined to be due to oxygen vacancies in the rutile structure of IrO$_2$. Additionally, the number density of these mobile defects can be calculated from the $\sqrt{T}$ resistance rise caused by the orbital two-channel Kondo effect in the Dirac nodal line metal IrO$_2$.
\end{abstract}

\maketitle
The metal iridium dioxide (IrO$_2$) has drawn significant attention as a promising material for spintronic applications due to its giant spin Hall resistivity.\cite{Fujiwara.2013} IrO$_2$ is also widely used in glucose and pH sensors.\cite{DONG2018} Recently, the elusive orbital two-channel Kondo (2CK) effect was observed in IrO$_2$ nanowires (NWs).\cite{Yeh2020} It results in a square-root-in-temperature increase of the resistivity, {\itshape i.e.,} $\rho \propto -\sqrt{T}$ (where $\rho$ is resistivity, and $T$ is temperature) below a characteristic temperature due to oxygen vacancies (V$_{\rm O}$'s).\cite{Yeh2020} These observations make IrO$_2$ NWs an attractive platform for exploring strongly correlated electron systems.\cite{Kirchner2020} Despite its potential, there is limited information on the low-$T$ low-frequency $1/f$ noise and its origin in IrO$_2$. The $1/f$ noise is a key factor which will ultimately limit the performance of every nanoelectronic device.\cite{Balandin2013} 

The standard theory of 1/$f$ noise in metals assumes the existence of mobile defects, often modeled as a set of independent two-level systems (TLSs) with a wide distribution of relaxation times. It can be shown that the resulting resistance fluctuations power spectrum density (PSD) possesses a $1/f^\alpha$ dependence, with $\alpha \simeq 1$.\cite{Dutta1981,Weissman1988,Fleetwood2015}
In this study, we examine the relationship between the number density of mobile defects ($n_m$) in IrO$_2$ NWs and the $1/f$ noise measured within the $T$ range of 1.7 to 350 K. We find unexpected results at low $T$, where the noise magnitude increases with decreasing $T$. We explain that this low-$T$ $1/f$ noise originates from the universal conductance fluctuations (UCF), which stem from quantum interference of conduction electrons scattered off mobile defects.\cite{Giordano1991,Feng1991,Gupta1994} We identify these  mobile defects are associate with the V$_{\rm O}$'s in the IrO$_2$ rutile structure. While 1/$f$ noise has been measured in a wide variety of conductors, both the origin and the amount of the underlying mobile defects in a given material are usually difficult to identify and  quantize. In the present case, we demonstrate that combined with the measurements of the 2CK effect, the $n_m$ value in every NW can be inferred.

\begin{figure}
	\centering
	\includegraphics[width=1\linewidth]{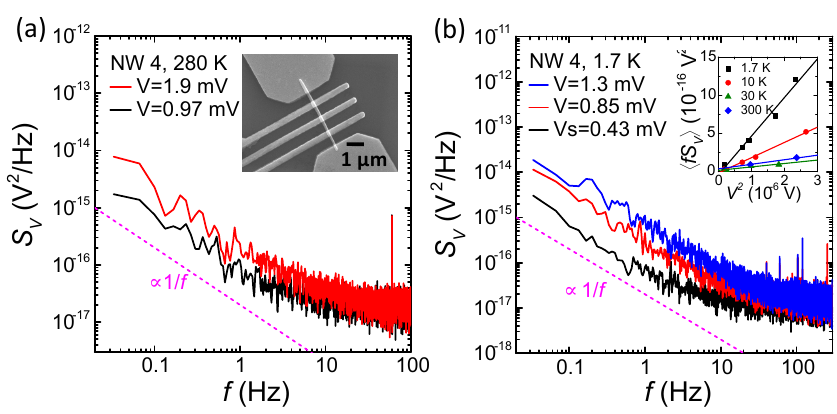}
	\caption{Voltage power spectrum density as a function of frequency for NW 4 at (a) 280 K and (b) 1.7 K. The dashed line is drawn proportional to $1/f$ and is a guide to the eye. Insets: (a) an SEM image of a typical IrO$_2$ NW device, and (b) $\langle f S_V \rangle$ versus $V^2$ at four $T$ values.  
	}
	\label{fig_1}
\end{figure}

Our IrO$_2$ NWs were grown via the metal-organic chemical vapor deposition method.\cite{Chen2004a,Lin2008} The single-crystalline rutile structure was characterized by selected-area electron diffraction patterns\cite{Chen2004a} and X-ray diffraction (XRD) patterns.\cite{Chen2004b} To generate V$_{\rm O}$'s, the NWs were thermally annealed at 300$^\circ$C in vacuum. The annealing time was 3 h for NWs 1 to 3, and 4 h for NW 4. To facilitate electrical measurements, submicron Cr/Au (10/120 nm) electrodes attaching a NW device were fabricated by the electron-beam lithographic technique. The inset of Fig. \ref{fig_1} shows a scanning electron microscopy (SEM) image of a typical NW device. An ac method for 1/$f$ noise measurement was applied.\cite{Scofield1987} The background voltage noise PSD in our setup was $S_V^0 \approx 1.6\times 10^{-17}$ V$^2$/Hz, which was limited by the voltage preamplifier (Stanford Research model SR560). The relevant parameters of the four IrO$_2$ NWs studied in the present work are listed in Table \ref{tableI}.

\begin{table}
\caption{\label{tableI}
Relevant parameters for IrO$_2$ NWs. $L$ is the NW length between the voltage probes in a 4-probe configuration, $d$ is the NW diameter, 
$n_m^{\rm 2CK}$ is number density of mobile defects determined from the 2CK effect, $n_i$ is number density of total (static and mobile) defects, $L_e$ is electron elastic mean free path, and $L_\varphi$ is electron dephasing length. $(L_e/L_\varphi)^2$ is calculated with $L_\varphi \simeq 100$ nm.
}
\begin{ruledtabular}
\begin{tabular}{cccccc}
Nanowire & $L$ ($\mu$m) & $d$ (nm) & $n_m^{\rm 2CK}$ (m$^{-3}$)  & $n_m^{\rm 2CK}/n_i$   & $(L_e/L_{\varphi})^2$ \\ 
\hline 
NW 1 & 1.9 & 143 & $7.5 \times 10^{25}$  & 0.074 & 0.0015 \\
NW 2 & 1.0 & 123 & $1.8 \times 10^{25}$  & 0.020 & 0.0019 \\
NW 3 & 0.75 & 125 & $3.4 \times 10^{25}$ & 0.037 & 0.0018 \\
NW 4 & 0.90 & 147 & $7.8 \times 10^{25}$ & 0.065 & 0.0010 \\
\end{tabular}
\end{ruledtabular}
\end{table}

\textit{Power spectrum density and Hooge parameter $\gamma$.} For an ohmic conductor under a small bias current $I$, the measured voltage noise is usually expressed by the empirical form\cite{Hooge1969} 
\begin{equation}
S_V=\frac{\gamma}{N_c f^{\alpha}}V^2+S_V^{0}\,,
\label{eq:Hooge}
\end{equation}
where $\gamma$ is the Hooge parameter which characterizes the 1/$f$ noise magnitude, $N_c$ is the total carrier number in the sample, $V$ is the voltage dropped on the sample.

Figures \ref{fig_1}(a) and (b) show the measured $S_V$ of NW 4 at two representative temperatures 280 and 1.7 K, respectively. Above several tens Hz, the measured $S_V$ approaches a constant ($\approx S_V^0$). At lower frequencies, $S_V$ increases with decreasing $f$, as well as with increasing $V$. A $S_V \propto f^{-\alpha}$ dependence, with $\alpha \simeq 1$, is found for $f \lesssim 1$ Hz. This dependence can be well described by Eq. (\ref{eq:Hooge}). To extract the $\gamma$ value, we rewrite Eq.~(\ref{eq:Hooge}) as $\left \langle fS_V \right \rangle = \left \langle \gamma V^2/N_c \right \rangle + \left \langle fS_V^0 \right \rangle = \gamma V^2/N_c + \left \langle f \right \rangle S_V^0 $, where $\left \langle fS_V \right \rangle$ denotes the average of the product of each discrete $f_i$ and $S_{Vi}$ in the data set in the $S_V \propto 1/f$ regime. The inset of Fig. \ref{fig_1}(b) shows that our data obey the $\left \langle fS_V \right \rangle \propto V^2$ dependence at our measurement temperatures. Thus, the $\gamma$ value can be obtained from a linear fit of the slope $\gamma/N_c$, where $N_c=n_c L d^{2}$, with $n_c \approx 1 \times 10^{28}$ m$^{-3}$ being the carrier density of IrO$_2$,\cite{Kawasaki2018} $L$ ($d$) the length (diameter) of the NW. The linear dependence $\left \langle fS_V \right \rangle \propto V^2$ suggests that the measured voltage noise originates from the resistance fluctuations in the NW, \textit{i.e.}, the (small) applied $I$ only acts as a sensitive electrical probe, while it does not drive the resistance fluctuations.\cite{Hooge1969} 

\begin{figure}
	\centering
	\includegraphics[width=1.0\linewidth]{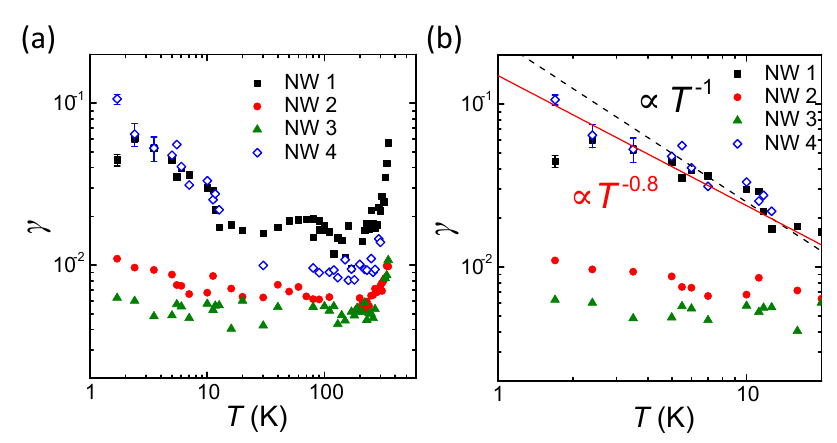}
	\caption{(a) Hooge parameter $\gamma$ as a function of temperature for four IrO$_2$ NWs between 1.7 and 350 K. Note that in NWs 1 and 4, $\gamma$(2\,K)$>$$\gamma$(300\,K). (b) $\gamma$ versus $T$ below 20 K. The straight solid line is a linear fit to NW 4. The straight dashed line is a guide to the eye. In NW 1, $\gamma$ saturates below about 3 K.
	}
	\label{fig_2}
\end{figure}

The extracted $\gamma$ values as a function of $T$ between 1.7 and 350 K for four IrO$_2$ NWs are plotted in Fig. 2(a). As $T$ decreases from 350 K, $\gamma$ first decreases and then saturates below around 100 K. For example, the $\gamma$ value of NW 4 decreases from $\approx$\,0.014 at 300 K to $\approx$\,0.009 at 100 K, and then remains constant until about 20 K. In this high-$T$ regime, we expect that the TLSs are thermally activated and the $1/f$ noise essentially follows the local-interference model.\cite{Pelz1987} In the rest of this Letter, we shall focus our discussion on the noise  below 20 K, which results from the UCF in the NWs.

\textit{UCF-induced 1/$f$ noise.} The most interesting observation of this work is that, below about 20 K, our extracted $\gamma$ values increase with decreasing $T$ in all NWs. This extraordinary behavior is particularly significant in NWs 1 and 4. In NW 4, the value of $\gamma$ at 1.7 K is  $\simeq 0.11$ [$\gamma(1.7\,{\rm K}) \simeq 0.11$] and is one order of magnitude larger than that at 30 K, and even larger than that at 300 K. Figure \ref{fig_2}(b) shows that NW 1 has $\gamma (T)$ behavior similar to that of NW 4, except that the $\gamma$ value saturates below $\sim$\,3 K. The low-$T$ increase of $\gamma$ in NWs 2 and 3 is smaller than that in NWs 1 and 4.

The rapid and repeated switching of mobile defects or TLSs modifies (a large fraction of) the electron paths traversing the NW. The resulting  quantum interference among the electron trajectories is sensitively altered. The interference effect becomes more pronounced at lower $T$, where the electron dephasing time ($\tau_{\varphi}$) well surpasses the electron elastic mean free time ($\tau_e$), \textit{i.e.}, $\tau_{\varphi} \gg \tau_e$. This is the origin of the temporal UCF.\cite{Feng1991,Giordano1991} In 1991, Feng realized that UCF in turn could give rise to 1/$f$ noise. He showed that the phenomenological Hooge parameter $\gamma$ was directly connected with the variance of the conductance fluctuations, and given by,\cite{Feng1991,Giordano1991}
\begin{equation}
\gamma= \frac{N_c}{{\rm ln} (f_{\rm max}/f_{\rm min})} \frac{\delta G^2} {G^2}\,,
\label{eq:gamma-deltaG}
\end{equation} 
where $f_{\rm max}$ ($f_{\rm min}$) denotes the maximum (minimum) frequency of the mobile defects which are relevant in the measurement, $G$ is the conductance of the sample, and $\delta G^2$ is the variance of the UCF. 
Because $\delta G^2/G^2$ increases as the quantum-interference effect increases, $\gamma$ grows with decreasing $T$. An additional $T$-dependence of $\delta G^2/G^2$ may be due to the saturation of $\tau_\varphi$ ({\itshape i.e.}, $\tau_\varphi$ becomes nearly independent of $T$) as $T \rightarrow 0$.\cite{Lin2002}

As the electron dephasing length $L_\varphi = \sqrt{D\tau_\varphi} < L$, $d$, where $D = v_F^2 \tau_e/3$ is the electron diffusion constant and $v_F$ is the Fermi velocity, our NWs are essentially three-dimensional (3D). We consider the UCF in the ``saturated'' regime, which is defined by $n_m/n_i \gg (L_e/L_{\varphi})^2$, where $n_i = 1/L_e \langle \sigma \rangle$ is the number density of total (static and mobile) defects, $L_e = v_F \tau_e$ is the electron elastic mean free path, and $\langle \sigma \rangle$ is the averaged electron scattering cross section. This is the regime that is pertinent to our NWs, see Table \ref{tableI}. The UCF theory predicts under these conditions the normalized variance\cite{Feng1991}
\begin{equation}
\frac{\delta G^2}{G^2}=28.2 \frac{1}{k_{F}^4 L_{e}^2} \frac{L_\varphi}{Ld^2}\,,
\label{eq:UCF}
\end{equation} 
where $k_F$ is the Fermi wavenumber. Thus, the $T$ dependence is fully determined by the $T$ behavior of $L_{\varphi}$. In writing down Eq. (\ref{eq:UCF}), the thermal energy averaging effect has been ignored which presumably would introduce another characteristic length scale, the thermal diffusion length $L_T = \sqrt{D\hbar/k_BT}$, into the cutoff of the UCF ($\hbar$ is the Planck constant divided by 2$\pi$, and $k_B$ is the Boltzmann constant). We have previously found that the thermal energy averaging effect on the UCF magnitude was negligibly smaller than theoretically predicted.\cite{Yang2012} Thus, we take Eq. (\ref{eq:UCF}) to be sufficient for providing a good understanding of the $\gamma (T)$ data.  

Figure \ref{fig_2}(b) shows that below about 20 K, $\gamma \propto L_\varphi \propto T^{-0.8}$ in NW 4. This $T$ dependence can be readily explained in terms of an electron dephasing rate $1/\tau_\varphi (T) = 1/\tau_\varphi^0 + 1/\tau_{e-ph}(T)$, where the first term $1/\tau_\varphi^0$ has been widely observed in experiments and is called the saturated electron dephasing rate as $T \rightarrow 0$.\cite{Lin2002} The second term $1/\tau_{e-ph} \propto T^2$ is the electron-phonon scattering rate which dominates the inelastic scattering in weakly disordered metals.\cite{Zhong2010,Zhong1998} In NW 1, $\gamma$ shows a similar $T$ behavior until it saturates below about 3 K, indicating that $1/\tau_\varphi^0$ is much larger in this NW than in NW 4.\cite{Trionfi2007} The magnitudes of $\gamma$ increase in NWs 2 and 3 are small, because these two NWs have comparatively small $n_m$ values (see below).

\begin{figure}
	\centering
	\includegraphics[width=1.0\linewidth]{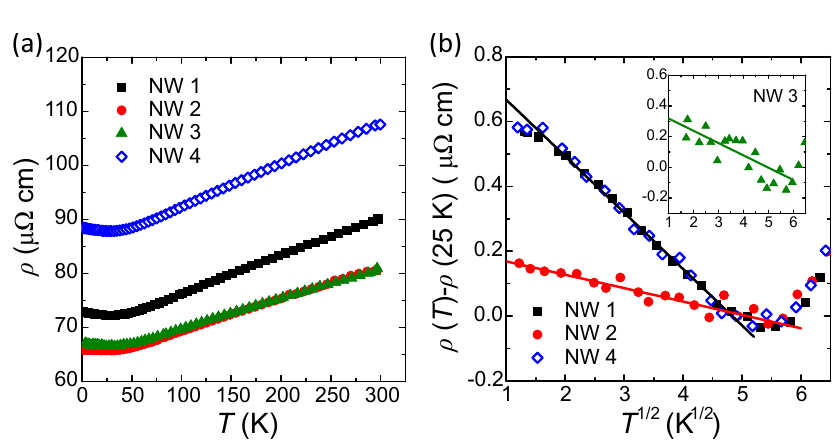}
	\caption{(a) Resistivity versus temperature for four IrO$_2$ NWs. The data reveal typical metallic behavior above $\sim$\,25 K. At lower $T$, $\rho$ increases with decreasing $T$. (b) Main panel and inset: $[\rho(T)-\rho (25\, {\rm K})]$ versus $\sqrt{T}$. The straight lines are linear fits to Eq. (\ref{eq:2CK}).
 	}
	\label{fig_3}
\end{figure}

\textit{Identifying the origin and number density of mobile defects.} 
We will now delve into the examination of the microscopic source of mobile defects responsible for inducing UCF and determining the 1/$f$ noise magnitude in IrO$_2$ NWs, including their number density. This is based on  nonmagnetic orbital 2CK physics which was recently reported in this material. The 2CK effect in IrO$_2$ causes a $\sqrt{T}$ increase of $\rho$ at low $T$. Figure \ref{fig_3}(a) shows the variation of $\rho$ with $T$ for our NWs. $\rho(T)$ reveals typical Boltzmann transport behavior until below about 25 K where it slightly increases with decreasing $T$. Close inspection indicates that this increase obeys a $\sqrt{T}$ dependence down to 2 K, Fig. \ref{fig_3}(b). Between 2 and 25 K, the normalized increase is $\left[ \rho \left( 2~ {\rm K} \right) -\rho \left( 25~ {\rm K} \right) \right] /\rho \left( 25~ {\rm K} \right) \approx (2.2 - 7.7) \times 10^{-3}$ in our NWs. This amount of resistivity increase is more than one order of magnitude larger than what would be expected from the 3D weak localization and electron-electron interaction effects.\cite{Lee1985,Yeh2020} We have recently demonstrated that this $\sqrt{T}$ increase obeys \cite{Affleck1993} 
\begin{equation}
    \rho_{\rm 2CK}\left( T \right) = \frac{3n_m^{\rm 2CK}}{4\pi \hbar \left[ eN(0)v_F \right]^2} \left( 1-4 \sqrt{\frac{\pi T}{T_{\rm 2CK}}} \right)\,,
    \label{eq:2CK}
\end{equation}
where $n_m^{\rm 2CK}$ denotes the number density of 2CK scatterers, $N \left( 0 \right)$ is the density of states at the Fermi energy per spin per channel, and $T_{\rm 2CK}$ is a characteristic temperature scale in the 2CK problem. 

Oxygen vacancies frequently occur in IrO$_2$ which assumes the rutile structure, depending on the sample preparation conditions.\cite{Yeh2020} Due to charge neutrality, a V$_{\rm O}$ generates two ``defect electrons'' in IrO$_2$. One of the two defect electrons, which occupies the doubly degenerate 5$d_{xz}$ or 5$d_{yz}$ iridium ion orbital, is strongly coupled to the two equivalent and independent spin-up and spin-down conduction-electron channels. This is the origin of the 2CK effect in IrO$_2$.\cite{Yeh2020,Kirchner2020} With the assumption that the density of V$_{\rm O}$'s, denoted by $n_{\rm V_O}$, is low enough to safely ignore any  correlations between V$_{\rm O}$'s, we can estimate $n_{\rm V_O} \simeq n_m^{\rm 2CK}$ in each IrO$_2$ NW.

To compare our data with Eq. (\ref{eq:2CK}), we take $T_{\rm 2CK} \simeq 23$ K, the temperature below which the $\sqrt{T}$ behavior is found. Then, the $n_m^{\rm 2CK}$ value in each NW can be extracted from the fitted slope plotted in Fig. \ref{fig_3}(b), using $N \left( 0 \right) \approx 1.5 \times 10^{47}$ J$^{-1}$ m$^{-3}$ and $v_F \approx 5 \times 10^5$ m/s for IrO$_2$.\cite{Lin1999,Yeh2020} These values are listed in Table \ref{tableI}. We see that NWs 1 and 4, which have the largest noise magnitudes $\gamma$, also show the largest $\sqrt{T}$ resistivity increases ({\itshape i.e.}, largest $n_m^{\rm 2CK}$ values), and NWs 2 and 3 have smaller noise magnitudes and simultaneously smaller resistivity increases ({\itshape i.e.}, smaller $n_m^{\rm 2CK}$ values). 

In practice, the mobile defects existing in a given sample may not all contribute to a given noise measurement, and $n_m$ will be given by the number of those defects having switching frequency $f_{\rm min} \leq f \leq f_{\rm max}$. Then, $n_m \lesssim n_{\rm V_O} \simeq n_m^{\rm 2CK}$. In the regime where the resistance fluctuations are governed by the local-interference (LI) mechanism and $n_{\rm V_O} \approx n_m^{\rm 2CK}$ holds, we obtain an estimate for the switching TLS density $n_m^{\rm LI}$. This is accomplished by applying a method developed earlier,\cite{Yeh2018} where $n_m^{\rm LI}$ can be estimated with the measured $\gamma$ value through $n_m^{\rm LI} \approx 4 \pi \gamma n_c \left( \rho e^2 / m v_F \left \langle \sigma \right \rangle \right)^2$, where $e$ ($m$) is the electronic charge (effective mass). At 30 K, this gives in the present case $n_m^{\rm LI} \approx (1.9 \pm 0.1) \times 10^{25}$ m$^{-3}$ for NWs 1 and 4, and $\approx (5.5 \pm 1) \times 10^{24}$ m$^{-3}$ for NWs 2 and 3. These extracted $n_m^{\rm LI}$ values scale with and are smaller than the $n_{\rm V_O}$ values within a factor of 10, indicating the fluctuators at low $T$ are associated with the V$_{\rm O}$'s. Thus, in our NWs, we estimate $n_m$ $[\sim n_m^{\rm LI}(30\,{\rm K}) \lesssim n_{\rm V_O} \simeq n_m^{\rm 2CK}]$ has values $\sim 10^{24}-10^{25}$ m$^{-3}$, being about one order of magnitude larger than that found in, {\itshape e.g.}, Bi films.\cite{Birge1989}

One possible explanation for this is the existence of an interstitial oxygen density that is (by and large) proportional to $n_{\rm V_O}$. Our thermal annealing studies suggest that this is not the reason for the proportionality between $n_{\rm V_O}$ and $n_m^{\rm LI}$, as an increase in annealing time in vacuum is expected to reduce the number density of oxygen interstituals while increasing $n_{\rm V_O}$. This leaves us with the more interesting alternative that the V$_{\rm O}$'s themselves are driving the 1/$f$ noise. 

The characteristic feature of the one-channel (two-channel) Kondo effect is the existence of a degenerate local level hybridizing with a band (two bands)
of electron states. This leads to a characteristic energy scale known as the Kondo temperature ($T_{\rm K}$) or equivalently a characteristic time scale $\tau_{\rm K}=\hbar/k_BT_{\rm K}$.\cite{Hewson1993} (In the present case, $T_{\rm K}$ = $T_{\rm 2CK}$.) On the other hand, 1/$f$ noise is associated with a broad distribution of relaxation times of the mobile defects. To ensure that the 2CK effect can form in IrO$_2$, the degeneracy between  5$d_{xz}$ and 5$d_{yz}$ has to remain intact even in the presence of such a broad distribution. Moreover, to guarantee that the flickering is not interfering with the formation of the 2CK effect, $k_BT_{\rm 2CK}\gg hf$ needs to hold. This is evidently satisfied, as can be inferred from Fig.\,\ref{fig_1} with $T_{\rm 2CK}\approx 20$ K as is appropriate for IrO$_2$ NWs. The necessary broad distribution of relaxation times of the fluctuators driving the 1/$f$ noise implies a corresponding disorder distribution of the V$_{\rm O}$ environment. That the $T_{\rm K}$ value distribution in such a situation remains sharply peaked has been addressed before.\cite{Chakravarty.00}
A microscopic model of how V$_{\rm O}$'s can drive an orbital Kondo effect and concomitantly 1/$f$ noise at lower frequencies is left for future work.
Intuitively, if different V$_{\rm O}$'s reside at varying distances to grain boundaries, and grain boundaries in the NW have differing orientations, a distribution of the relaxation times should naturally arise. NW surfaces (and the interface with substrate) could likely have similar effects on resulting in a distribution of  relaxation times. In any case, our recent experiments have confirmed that the 1/$f$ noise magnitude in a series of RuO$_2$ films is controlled by $n_{\rm V_O}$, which can be reversely adjusted, {\itshape i.e.}, increased or reduced, by annealing RuO$_2$ in vacuum, air, or  oxygen gas.\cite{Yeh2018} Since both RuO$_2$ and IrO$_2$ crystallize in the rutile structure and have similar physical properties\cite{Lin2004} (except that RuO$_2$ is antiferromagnetic which leads to an orbital one-channel Kondo effect at low $T$, see Ref. \onlinecite{Yeh2020}), we expect that V$_{\rm O}$'s play a similar role in producing 1/$f$ noise in IrO$_2$.

{\itshape Temporal resistance fluctuations.} While the UCF-induced 1/$f$ noise has been theoretically proposed in the 1980s,\cite{Feng1986} the predictions have not been widely tested. The reasons are that one needs to use microscale samples and measure the noise down to low $T$. Otherwise, the quantum-interference effect will be negligible and the noise magnitude small and difficult to detect. To further check that the measured 1/$f$ noise in IrO$_2$ below 20 K is caused by the UCF mechanism, we have extracted the (normalized) variance of conductance fluctuations $\delta G^2 / G^2$ from an independent measurement, and compare it with that calculated from the measured $\gamma$ value through Eq. (\ref{fig_2}).

\begin{figure}
	\centering
	\includegraphics[width=1.0\linewidth]{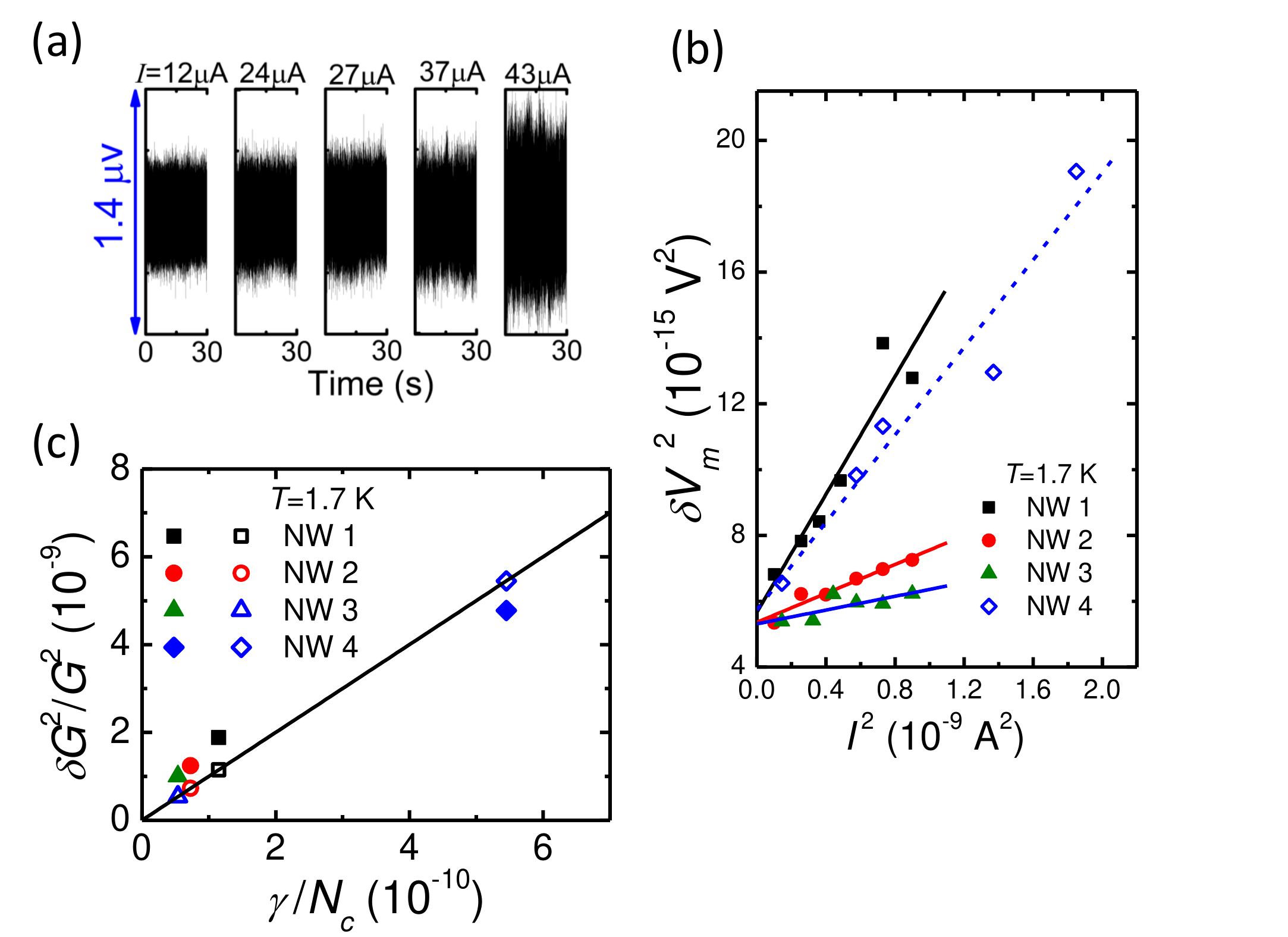}
    \caption{(a) Voltage fluctuations as a function of time for NW 4 at 1.7 K and under five different bias currents. (b) $\delta V_m^2$ versus $I^2$. Straight solid and dashed lines are linear fits. (c) $\delta G^2/G^2$ versus $\gamma /N_c$. Closed symbols are calculated from $\delta V_m^2$. Open symbols are calculated through Eq. (\ref{eq:gamma-deltaG}). The straight line is a guide to the eye.
	}
	\label{fig_4}
\end{figure}

In our experimental method, a current $I$ is applied to the NW device and the total voltage fluctuations $\delta V_m (t)$ are measured, {\itshape i.e.}, $\delta V_m (t)=I \delta R(t)+\delta V_0 (t)$, where $I \delta R$ is the voltage noise of the NW, and $\delta V_0$ is the input noise of the preamplifier. Figure \ref{fig_4}(a) shows the measured $\delta V_m (t)$ at several $I$ values for NW 4 at 1.7 K. We see that the $\delta V_m$ magnitude increases with increasing $I$. The $\delta G^2 / G^2$ value of a NW can be obtained by subtracting out the background contribution $\delta V_0$. Because $\delta R(t)$ and $\delta V_0 (t)$ are uncorrelated, we write $\delta V_m^2 = I^2 \delta R^2 + \delta V_0^2$ for an ohmic conductor, where $\delta V_m^2$, $\delta R^2$, and $\delta V_0^2$ denote the variances of $\delta V_m (t)$, $\delta R(t)$, and $\delta V_0 (t)$, respectively. Thus, the intrinsic resistance fluctuations of the NW $\delta R^2$ can be obtained from a plot of $\delta V_m^2$ versus $I^2$. Figure \ref{fig_4}(b) shows the variation of $\delta V_m^2$ with $I^2$ for our NWs at 1.7 K. A $\delta V_m^2 \propto I^2$ linear dependence is observed for all NWs, with the fitted $\delta V_0^2 \approx (5.5 \pm 0.2) \times 10^{-15}$ V$^2$. Here we note that in our noise measurements, the sampling rate of the spectrum analyzer (Stanford Research model SR785) was set to be 1024 Hz and the measurement time was lasted for 30 s, corresponding to the relevant frequency range from $f_{\rm min} \simeq 0.033$ Hz to $f_{\rm max} \simeq 512$ Hz. This $f$ range leads to an estimated upper bound $\delta V_0^2 \sim S_V^0 \times \left( f_{\rm max} - f_{\rm min} \right) \sim 8\times 10^{-15}$ V$^{2}$, being in good accord with our measurement.

Once $\delta R^2$ value in each NW is obtained from the slope in Fig. \ref{fig_4}(b), $\delta G^2 / G^2 = \delta R^2 / R^2$ can be calculated. The calculated $\delta G^2 / G^2$ values (solid symbols) as a function of the normalized Hooge parameter $\gamma/N_c$ are plotted in Fig. \ref{fig_4}(c). Alternatively, we have calculated $\delta G^2 / G^2$ values (open symbols) from the measured $\gamma$ (Fig. \ref{fig_2}) through Eq. (\ref{eq:gamma-deltaG}), by inserting ${\rm ln} (f_{\rm max}/f_{\rm min}) \approx {\rm ln} (512/0.033) \approx 10$. Figure \ref{fig_4}(c) shows that the $\delta G^2 / G^2$ values obtained from the two independent methods agree to within a factor of $\sim$\,2. At $T$ = 1.7 K, $\delta G^2 / G^2 \approx 1.9 \times 10^{-9}$ and $\approx 4.8 \times 10^{-9}$ in NWs 1 and 4, respectively. We thus obtain $L_{\varphi} \approx$ 95 (81) nm in the former (the latter) through Eq. (\ref{eq:UCF}), where $k_F \approx 7 \times 10^9$ m$^{-1}$.\cite{Lin1999} Therefore, we have $L_{\varphi} < d$ and $n_m/n_i \gg (L_e/L_{\varphi})^2$, justifying the application of Eq. (\ref{eq:UCF}). 

The 1/$f$ noise has previously been studied in a wide variety of metals\cite{Dutta1981,Weissman1988,Giordano1991} and semiconductors,\cite{Fleetwood2015} and recently, in 2D materials.\cite{Balandin2013,Karnatak2017} UCF-induced 1/$f$ noise were observed in polycrystalline Bi films,\cite{Birge1989} Ag films,\cite{Meisenheimer1989,Birge1993} Au wires,\cite{Trionfi2007} RuO$_2$ NWs,\cite{Lien2011} {C-Cu composites,\cite{Garfunkel1988} and epitaxial graphene.\cite{Kalmbach2016} In Bi and Ag films, the UCF fell in the ``unsaturated'' regime where $n_m/n_i \ll (L_e/L_\varphi)^2$. The origins of the mobile defects were unspecified in those samples, while the diffusion of H atoms (the spin-glass freezing process) was identified as the cause in amorphous Ni-Zr films\cite{Alers1989} (narrow AuFe wires\cite{Neuttiens2000}). In the present study, we are able to show that V$_{\rm O}$'s are rich and generate mobile defects in IrO$_2$. 

In summary, we have measured the 1/$f$ noise of four IrO$_2$ nanowires in the temperature range from 1.7 to 350 K. As temperature decreases from 350 K, the $\gamma$ value decreases down to about 100 K and then remains constant, but as $T \lesssim 20$ K $\gamma$ increases with decreasing temperature. We explain this low-temperature $\gamma$ increase in terms of the UCF-induced 1/$f$ noise. We identify the mobile defects to be associated with oxygen vacancies in the IrO$_2$ rutile structure. In combination with the orbital 2CK effect previously observed in this Dirac nodal line metal, we have extracted the number density of the mobile defects. The UCF origin of the 1/$f$ noise is further confirmed by independent measurements of temporal conductance fluctuations in the nanowires. 
\\
\\
\noindent\textbf{ACKNOWLEDGMENTS}\\

We thank C. W. Wu for experimental help. This work was supported by National Science and Technology Council of Taiwan through grant numbers 110-2112-M-A49-015 and 111-2119-M-007-005 (J.J.L.), and 110-2112-M-A49-033-MY3 (S.S.Y.). \\ 







\noindent\textbf{REFERENCES}

%

\end{document}